\documentclass[preprintnumbers,amsmath,amssymb]{revtex4}
\usepackage{graphicx}
\usepackage{graphics,epsfig,psfig}
\usepackage{bm}
\usepackage{amssymb}

\begin{document}


\title{ A deductive statistical mechanics approach for granular matter }


\author{ T. Aste and T. Di Matteo}


\address{ Department of Applied Mathematics, \\
The Australian National University, 0200 Canberra, ACT, Australia. }


\begin{abstract}
We introduce a deductive statistical mechanics approach for granular materials which is formally built from few realistic physical assumptions. 
The main finding is an universal behavior for the distribution of the density fluctuations.
Such a distribution is the equivalent of the Maxwell-Boltzmann's distribution in the kinetic theory of gasses.
The comparison with a very extensive set of experimental and  simulation data for packings of monosized spherical grains, reveals a remarkably good quantitative agreement with the theoretical predictions for the density fluctuations both at the grain level and at the global system level. 
Such agreement is robust over a broad range of packing fractions and it is observed in several distinct systems prepared by using different methods.
The equilibrium distributions are characterized by only one parameter ($k$) which is a quantity very sensitive to changes in the structural organization.
The thermodynamical equivalent of $k$ and its relation with the `granular temperature' are also discussed.
\end{abstract}

\maketitle

\section{Introduction}

Granular  materials are central in a very wide range of domains from agriculture to pharmaceutical industry, but, despite such a central role in most fields of human activity and their ubiquitous presence in scientific research areas, their behaviour and properties remain elusive. 
Granular materials differ from all the other materials: they can flow like liquids under some conditions and act as solids under others 
\cite{Umbanhowar03,Buchanan03,Hecke05}. 
Their intrinsic complexity is consequence of the fact that they are composed of many pieces (the grains) that can assemble in large structures behaving collectively. 
The study of granular materials has forced scientists to rethink established classifications of matter and to reformulate statistical mechanics in a new context.
These studies are springing out new ideas that are advancing the understanding of large classes of complex materials from composites to foams. 
 
The question whether a statistical mechanics approach can correctly describe granular systems has sparkled an intense debate in the scientific community in recent years 
\cite{Edwards89,Barrat00,Fierro02,Makse02,Behringer02,DAnna03,Srebro03,Blumenfeld03,Edwards04,Ojha04,Snoeijer04,Richard05,Schroder05,Blumenfeld06,ChamarraPRL06,Lechenault06,Metzger04,Metzger05,Gao06}.
Such a statistical mechanics theory of granular matter must be able to connect the observable behavior of large aggregates with the structural and dynamical properties of the grains constituting these systems.
The great interest in such `thermodynamical' theory of granular packings resides in the fact that it would lead to a description of such a complex system in terms of a few relevant parameters only \cite{Makse02,Behringer02,DAnna03,Ojha04,Richard05}.  
This is in analogy with traditional thermodynamics, where the state of a large system comprised of many interacting elements can be completely described by knowing temperature, volume or pressure.

In this paper we show that such approach is formally possible: a statistical mechanics formulation for granular matter can be built deductively from a small set of physical assumptions and the equivalent of the Maxwell-Boltzmann's distribution can be formally derived. 
More significantly, we show that such a distribution describes very accurately empirical and numerical data concerning volume distributions in sphere packings prepared with several different methods in a broad range of packing fractions.

One of the characteristics that differentiate granular materials from molecular materials is that frictional forces among grains can dissipate energy and drive the system towards frozen, jammed configurations. 
Because of such energy dissipation, granular materials can change their static configurations only when energy is injected in the system. 
This can be done in several ways, for instance by tapping the system with vertical vibrations or by pulsing fluids flows in fluidised beds or  by shearing or by rotating the container. 
The relevant fact is that such action should make possible for the system to change configuration and explore the accessible phase-space.

The statistical mechanics approach proposed in this paper is built within the same framework of the deductive foundation of statistical mechanics proposed by O. Penrose in \cite{PenroseBOOK70}.
The present paper is structured as follows.
In Section \ref{s.2}, we introduce the  general idea of the system's observational states and the probability of  transition between states when the system is driven by trains of energy injections. 
In Section \ref{s.1}, we define a microscopic `fine grained' description of system's state concerning grain packs at static equilibrium.
In this section we also discuss other possible `coarse grained' macroscopic descriptions and the links between these different descriptions.
A classification of states in terms of their transitional probabilities is proposed in Section \ref{s.3} where the concept of ergodic set is also introduced.
In Section \ref{s.3b}, we discuss the statistical equilibrium.
The system entropy is defined in Section \ref{s.4} where a maximum-entropy prediction for the probability of a given (coarse-grained) state is also introduced. 
The readers that are only interested on the theoretical outcomes might skip all the previous sections and jump directly to Section \ref{s.5} where the probability density function for the system's volume fluctuations is calculated.
In Section \ref{s.8}, the concept of `granular temperature' and the links with volume fluctuations at equilibrium are discussed. 
In Section \ref{s.6}, it is shown that the theoretical predictions for the  probability density function for local volumes at the grain levels (Vorono\"{\i} cells) reproduce extremely well experimental and simulation data.
Moreover, we show that the theory predicts also the global fluctuations on the scale of the whole system.
The signatures of  transitions occurring at the random loose packing limit and at the random close packing limit are highlighted by studying the changes occurring at these densities in a quantity ($k$) which is shown to be the analogous to the specific heat for these systems. 
Conclusions and perspectives are given in Section \ref{s.9}.

\section{Observational probability and transition between states}
\label{s.2}

Let us here consider a granular system that evolves by means of repeated energy injections.
We are interested in the system's state observed in between trains of energy injections when all the grains are at rest  in a stable static configuration.
Let us use the term `trials' to indicate such energy injection trains.
Let us define $P_t(\Psi \leftarrow \Psi_0)$ as the probability to find the system in the observational state $\Psi$ after an experiment consisting of $t$ trials, given that at the beginning of the experiment  ($t=0$) the system was prepared in the state $\Psi_0$. 
We will return on the definition of the  observational state in the next section. 
In this section we  assume that the system static state can be defined and encoded.

A further important assumption is  that the system is \emph{Markovian} \cite{PenroseBOOK70}: 
the current observational state embodies all the observational information about the past history of the system that are relevant to its future observational state.
Such assumption allows us to write the Chapman-Kolmogorov equation describing the evolution under energy injections  \cite{PenroseBOOK70}:
$ P_{t+1}(\Psi  \leftarrow \Psi_0) = \sum_{\Psi_1} P_{1}(\Psi  \leftarrow \Psi_1) P_{t}(\Psi_1  \leftarrow \Psi_0)$; which can be re-written in the more familiar form
\begin{equation}
p(\Psi,t+1) = \sum_{\Psi_1} w(\Psi \leftarrow \Psi_1) p(\Psi_1,t) \;\;;
\label{e.CK1}
\end{equation}
where $w(\Psi \leftarrow \Psi_1) = P_{1}(\Psi  \leftarrow \Psi_1) $ is the probability of transition between the state  $ \Psi_1$ to the state  $ \Psi $ by performing one trial on the system.
Whereas, $p(\Psi_1,t)$ is a simplified notation for  $P_{t}(\Psi_1,\Psi_0)$ with the initial state omitted.

The Markovian assumption is quite strong from a statistical point of view.
However, in the context we are investigating, this assuption is adequate.
Indeed, a granular system at rest in a static stable configuration can stay `frozen' in such a state for ever.
Such a state might depend on all the system past history, the Markovian hypothesis simply assumes that all the relevant information which will determine the system future evolution is contained in the properties of the present system's state.
Indeed, there are no simple mechanisms to store `hidden' information about the granular system which are not embodied into the \emph{complete} microscopic description of the system at rest. 
It must be noted that, on the contrary, a coarse-grained description of the system's state might classify as identical states with different microscopic properties.
Although belonging to the same coarse-grained classification, such states are distinct at microscopic level and might evolve with different transition probabilities.

\section{Definition of the system's state}
\label{s.1}

The aim of this section is to  show with a practical example that it is formally possible to precisely encode the system's microscopic state. 
Granular materials are made of several grains which are typically confined in a container.
Each grain `$i$' in the system can have a different shape ($ \mathbf s_i$) and a different mass ($m_i$).
The geometrical disposition of each grain can be characterized by the coordinates of its barycenter $\mathbf r_i  = (x_i, y_i, z_i)$ and the grain orientation $\mathbf o_i = (\theta_i,\phi_i)$.
If we assume that:
1) the grains are non-deformable objects but they can overlap; 
2) the grains interact only when they are in contact or they are overlapping; 
3) the interaction energy is a known function of the grain overlaps; 
then, it is possible to give a complete description of the observable state of a packing of $N$ grains at rest in a stable static configuration  by providing the set of positions and orientations of all grains: 
$\Psi = \{\mathbf  r_1, \mathbf r_2, ... , \mathbf r_N,  \mathbf o_1, \mathbf o_2, ... , \mathbf o_N \}$.

However, when friction is present, the tangential frictional force between two grains is a function of the relative path that has produced such static configuration \cite{Rubinstein06}. 
Such dependence on the previous history would falsify the Markovian assumption if we adopt the  previous classification for $\Psi $.
On the other hand, we can solve this apparent difficulty by including the set of tangential forces between grains into the microscopic description.
In this way all the information about the relevant past history is encoded in the preset state description. 
The future evolution depends only on the present state and the trials. 
The theory satisfies therefore the Markovian assumption.
Such an encoding is for instance used in numerical simulations of granular systems   such as the  Discrete Element Method (DEM)  with `soft' spheres \cite{Cundall79,Schafer96,Makse04}.

Independently on the details of the system and on its specific encoding, the previous example has shown a very important fact: the system microscopic state can be completely encoded in terms of a finite set of quantities. 
However, we must consider that there are several possible descriptions of the system's state and the use of different variables might be more appropriate in different systems.
For instance, it can be convenient to include in the local description of the packing also some information about the surrounding configuration.
To this purpose one can subdivide the packing into a set of $k$ local `elementary cells'.
Such cells can be viewed as the portions of space associated to local configurations in the  granular pack.
If we call $\mathbf c$ the properties of such cells (size, shape, volume, connectivity, center of mass, orientation, etc.), then the system's state can be conveniently classified in terms of the properties of such local `elementary cells': 
$\Psi = \{\mathbf  c_1, \mathbf c_2, ... , \mathbf c_k \}$. 
Let us note that the two descriptions in terms of positions and orientations or in terms of local cells must be equivalent and interchangeable. 
The complete information about the system's configuration is encoded in both the representations.
In the following we will refer to the state $\Psi$ as a `microscopic' state.

The classification  in terms of the microscopic state $\Psi$ is the most complete description of the system's static configuration.
However, in most of the practical cases such a `fine grained' description is neither necessary nor useful.
 There are indeed several different microscopic configurations that share the same macroscopic physical properties.
 A `coarse grained' description of the system's state can be therefore more useful, and practical, when one wants to associate structural properties to physical properties. 
An example of a macroscopic -- coarse grained -- description of the system's state is its classification in terms of the total volume $V$ occupied by the system (or equivalently in terms of the packing fraction, {\em volume of the grains} / V).
Certainly, there is a very large number of microscopic states $\Psi$ which correspond to the same volume $V$.
Nonetheless, there are several structural and physical properties of these systems that can be unambiguously classified in terms of the packing fraction only \cite{AstePRE05}. 
The possibility to construct a coherent framework which allows one to consistently describe the system at different levels of detail is one of the main features of statistical mechanics \cite{PenroseBOOK70}.
A statistical mechanics theory of granular material should provide the instruments to unambiguously describe the average system properties and their probability distributions at any chosen level of description.

\section{Classification of system's states}
\label{s.3}

The system's states can be classified in terms of their transitional probabilities \cite{PenroseBOOK70}.
In particular, a state $\Phi$ is \emph{transient} if there exists a state $\Psi$ that the system can reach from $\Phi$ but  it cannot go back to  $\Phi$. 
If the state is not transient it is said to be \emph{persistent}.
The set  $Z(\Phi)$ of all states that can be reached from a persistent state $\Phi$ is called \emph{ergodic set}. 
Persistent states may be separated into one or more distinct ergodic sets.
Once the system reaches a persistent state it will remain in the ergodic set containing such state and it will repeatedly visit  all the states in the set never going outside.
For large $t$ the system will eventually reach the limit probability distribution $p(\Psi,t\rightarrow\infty)= p_\infty(\Psi)$, that is the solution of the system of equations $p_\infty(\Phi) = \sum_{\Psi} w(\Phi \leftarrow \Psi) p_\infty(\Psi)$. 
Note that such limiting probability is \emph{stationary} in the sense that it is not dependent on $t$.
A system which is confined in an ergodic set $Z$ and  visits the states $\Psi\in Z$ with frequencies proportional to  $p_\infty(\Psi)$ is said to be at \emph{statistical equilibrium}.

We note that such transitional probabilities, which define the classification of states and ultimately the system's statistical equilibrium, are intrinsically dependent on the kind of trials we are performing on the system.
Therefore, the equilibrium state and even the ergodic set are not solely associated with the kind of system in exam.
Conversely, they are uniquely associated to the combination of both system properties (grain shapes, roughness, friction, container shape, ... ) and the kind of trial (tapping, fluid flow, pouring, rotating, ...) and its properties (vibration frequency and amplitude, fluid intensity, pouring height, rotation velocity,...).

These classifications of the system's states and all the transition properties  and evolution relations (Eq. \ref{e.CK1}) discussed in this section and in the previous one, can be applied also to coarse grained states, providing that they satisfy the Markovian assumption.
It is of some interest to note that a coarse grained  state $A$ can be at the statistical equilibrium even if the microscopic states haven't  reached equilibrium yet.

\section{Statistical equilibrium}
\label{s.3b}
We have already introduced in the previous section the definition of  \emph{statistical equilibrium}. 
This is a very important concept that must be discussed with some care.
In a molecular gas, statistical equilibrium is reached after a certain time when the system has achieved a homogeneous internal temperature and it has thermalized to the surrounding environment.
In the context of granular materials and in particular in tapping experiments, the `equilibrium' state is typically considered reached when, after a certain number of tappings with a given intensity, the packing fraction begins to oscillate around a constant value that is independent from the starting state.
Such an heuristic definition is consistent with the formal one given in the previous section, however there can be cases when more than one ergodic set can be reached from the same starting configuration. 
Each one of such ergodic sets can have different statistical equilibrium properties.

It must be noted that the definition of statistical equilibrium is intrinsically associated with the transitional probability and therefore with the kind of trial that we are repeatedly performing on the system, it is not a property of the system's static configuration.
In this sense it is misleading to think that statistical equilibrium can be reached by means of certain prescribed operations only (such as tapping or fluid flows).
On the contrary there are a very large number of actions (trials) that will drive the system toward equilibrium.
For instance, if we pour grains from a vessel into another and we do it several times being careful to repeat the operation each time in the same way (same height, inclination angle,...) we will eventually reach a situation in which the  static packing configurations have stable statistical properties.
Analogously, statistical equilibrium  can be achieved by rotating the container or by blowing air from below or by stirring the aggregate or by several other possible manipulations.
The open question is whether these `equilibrium' states are the same states or not.
They might have the same average packing fraction but, to coincide they must also belong to the same ergodic set. 
Generally speaking, it is quite unlikely that exactly the same regions of the phase-space are explored by means of operations as different as fluidized beds or pouring.
However, we will show in Section \ref{s.7} that there are evidences of remarkable universal properties that hold across very diverse preparation methods and different numerical simulations.
Such universality is  suggesting that there might be a very large class of processes that leads to the same ergodic set or to sets that are mostly overlapping. 

\section{Maximum Entropy prediction for the state probabilities}
 \label{s.maxent}
\label{s.4}

There are several different definitions of Entropy that can be applied to different contexts such as thermodynamics of molecular gasses or information theory.
In this paper we will use the `statistical entropy'  (or Gibbs entropy) $S(Z)$ which is better suited to be applied to the study of the ensemble of states $Z$ constituting the ergodic set:
\begin{equation}
S(Z)= - \sum_{A \in Z} p_\infty(A) \log p_\infty(A) +  \sum_{A \in Z} p_\infty(A) S(A) \;\;\;.
\label{e.SZ}
\end{equation}
Here the state $A$ is a given `coarse grained' description  of the system; $ p_\infty(A)$ is its --~stationary~-- probability within the ergodic set $Z$ and $S(A)$ is the entropy of the state $A$, which is:
$ S(A)= -\sum_{\Psi\in A}  q(\Psi) \log q(\Psi)$, where $q(\Psi)$ are the a-priori probabilities to find the system in the microscopic state $\Psi \in A$.
When all the microscopic states are a-priori equi-probable, the expression for $S(A)$ becomes
\begin{equation}
S(A)= \log \Omega(A) \;\;\;,
\label{e.SA}
\end{equation}
with $\Omega(A)$ the number of microscopic states which are classifiable under the same coarse grained state $A$.
Eqs.\ref{e.SZ} and \ref{e.SA} give a coherent definition of the system's entropy independently on the level of details at which the system's states are described. 
In general, the choice for the tuning of the system's description depends on our experimental capability of measuring the structural properties, and on our analytical or computational ability to evaluate the number of microscopic configurations which are contributing to such states.
For what concerns the a-priori probabilities $ q(\Psi)$, in the following we will assume that,  a-priori, the probabilities to find the system in any given microscopic state are all equal and therefore we will use Eq.\ref{e.SA}.
This is not an essential assumption for the present theory, however this is the most likely scenario for a granular system prepared by means of some random mixing operation.

Let us adopt  characterize the system's state in terms only of its total occupied volume $V$ (i.e. $A=V$).
Different experiments can result in granular packings with different total occupied volume $V$. 
For instance, in the classical experiment by \cite{Knight95,Nowak97,Nowak98}, different average packing fractions are obtained by driving the system with different tapping intensities. 
Here we are interested in the properties of the ergodic set, which is the set of all the possible total volumes which can be reached by means of the chosen system's handling.
The probability of each of such realizations is $p_\infty(V)$.
The ergodic set at statistical equilibrium is characterized by the average volume $\bar V$ (or equivalently the average packing fraction $\bar \rho$) and the study of the statistical properties of the ergodic set coincides with the study of the volume fluctuations around such average values.

The most important property of entropy is that it is a non decreasing function, therefore at the statistical equilibrium the entropy must reach its maximum. 
We can calculate $p_\infty(V)$ by searching for the functional form of the probability distribution function which maximizes the entropy  $S(Z)$ (Eq.\ref{e.SZ}).
Such maximization must be done under the condition that  the average  occupied volume is equal to $\bar V $, which is the characteristic average volume associated to the ergodic set $Z$ corresponding to the specific trial.
This yields to:
\begin{equation}
p_\infty(V) = \frac {  \Omega(V) e^{-   V/\chi   }   }{  \sum_{V'} \Omega(V') e^{- V '/\chi   } }\;\;\;,
\label{e.ME}
\end{equation}
with $ \chi^{-1}$ a Lagrange multiplier fixed by the constraint on the average volume: 
\begin{equation} \label{e.VmeL}
\bar V = \left< V \right> =  \sum_{V} V p_\infty(V)  \;\;.
\end{equation}
Eq.\ref{e.ME} is the --~almost unavoidable~-- outcome of any statistical mechanics theory.
The challenge is now to compute the number of microscopic states $ \Omega(V) $ associated  with coarse-grained states which occupy a total volume $V$. 
To this end we must introduce some further assumptions as discussed in the next section.

\section{Space partitions}
\label{s.5}

The phase-space volume  $\Omega(A)$ is the number of microscopic states which are classifiable under the same coarse grained state $A$.
Specifically, $\Omega(V)$ is associated with all the microscopic configurations  $\Phi = \{\mathbf  c_1, \mathbf c_2, ... , \mathbf c_k \}$ occupying a total volume $V$.
Let us assume that all the cell-properties ${\mathbf c}_i$ are either completely determined by their volumes $v_i$ or they are independent from  $v_i$.
Let us also assume that these cells can have arbitrary volumes above the minimum value  $v_{min}$, under the sole condition that the whole system must occupy a total volume $V$.
In this case, the volume of the accessible phase-space  can be computed exactly:
\begin{equation}
\Omega(V) = \frac{1}{\Lambda^{3k}} \int_{v_{min}}^V dv_1 \int_{v_{min}}^V dv_2 .... 
 \int_{v_{min}}^V dv_k \delta(v_1+v_2+...+v_k - V) 
 =   \frac{(V- k v_{min} )^{k-1} }{ \Lambda^{3k} (k-1)!} \;\;\;,
\label{e.S1Dbb}
\end{equation}
with $\Lambda$ a constant analogous to the Debye length.
Substituting into Eq.\ref{e.ME}, and by using Eq.\ref{e.VmeL} we obtain $\chi =(\bar V - k v_{min})/k$ and  the probability 
\begin{equation}
p_\infty(V) = f(V,k) = 
 \frac{k^{k}}{\Gamma(k) } \frac{(V - V_{min})^{(k-1)} }{(\bar V - V_{min})^{k}} \exp \left( {-k \frac{V-V_{min}}{\bar V - V_{min} } } \right)\;\;\;,
\label{e.pV1D}
\end{equation}
with $V_{min} = k v_{min}$.
The function $f(V,k)$ is the probability density function to find a packing of $k$ elementary cells occupying a volume $V$, in a set of experiments where the average occupied volume is $\bar V$.
In a previous paper we reported a different derivation of $f(V,k)$ from simple statistical arguments \cite{AsteEPL07}.
Note that Eq.\ref{e.S1Dbb} is valid for any $k$ and it holds even in the limit $k=1$.
Indeed, the observable system can be  any arbitrary  sub-set of a larger system.
Moreover, the experiment can be performed either on several different independent systems or --~equivalently~-- on several non-interacting sub-sets of a large system.
Eq.\ref{e.pV1D} is a Gamma distribution in the variable $V- V_{min}$; it is characterized by a `shape'  parameter  $k$ and a `scale'  parameter $(\bar V - V_{min})/k$ \cite{Gamma}. 
In the literature empirical fits with gamma distributions have been previously reported for two-dimensional Vorono\"{\i} networks \cite{Weaire86}, and for Vorono\"{\i} partitions from random Poisson points in three dimensions \cite{Pineda04}. 
Interestingly the same distribution has been observed in statistical study of crumpled paper \cite{Sultan06}.

The average volume from the distribution $f(V,k)$ coincides with  $ \bar V$ and the variance is
\begin{equation}
\sigma^2_v = \frac{(\bar V-V_{min})^2}{k} \;\;\;,
\label{e.variance}
\end{equation}
which is a useful relation to empirically evaluate $k$ from a set of volume measurements: $k =  (\bar V  -V_{min})^2/\sigma^2_v$.
We will make use of this relation in Section \ref{s.7}.

\section{The `degrees' of space partition}
\label{s.8}

Following Edward's ideas \cite{Edwards89,Mehta89}, a `granular temperature' can be inferred from an analogy with the thermodynamical relation $\beta =1/(k_BT) = \partial (Entropy)/\partial (Energy)$ \cite{Edwards89,Mehta89}, by susbstituting the volume to the role played by the energy in thermodynamical systems.
In the formalism of this paper this can be written as: $ \beta_{gr} = \chi^{-1} = \partial S(Z)/\partial \bar V $.
 From Equations \ref{e.SZ} and \ref{e.S1Dbb} we have:
 \begin{equation}
\beta_{gr}^{-1} =  \chi=  \frac{ \bar V  - V_{min}}{k} \;\;\;,
\label{e.granT}
\end{equation}
therefore the Edwards' compactivity  $\chi$ \cite{Edwards89,Mehta89} is the average free-volume per elementary cell, implying that the `granular temperature' is a measure of the kind and the degree of space-partition into elementary cells.
For a given system at a given packing fraction, the `granular temperature' is univocally associated to the quantity $k$.
By using Eqs.\ref{e.variance} and \ref{e.granT} we have
\begin{equation}
\chi=  \frac{\sigma_v^2}{ \bar V  - V_{min}}\;\;\;.
\label{e.EdwComapct}
\end{equation}
The present theory provides therefore a very effective and practical way to compute Edward's compactivity  from measures of volume fluctuations.

In terms of the compactivity $\chi$ the expression for the cell-volumes' distribution (Eq.\ref{e.pV1D} ) can be written in the more compact form
 \begin{equation}
p_{\infty}(V) = \frac{\Omega(V)}{\Omega(\chi)}e^{-\frac{V}{\chi}} \;\;\;,
\label{e.pinf}
\end{equation}
which recalls the equivalent expression in classical thermodynamics, with $\Omega(V)$ given by Eq.\ref{e.S1Dbb} and $\Omega(\chi) = \Omega_0 \chi^k \exp(-V_{min}/\chi)$.
The volume fluctuations within the ergodic set can be directly calculated from Eq.\ref{e.pV1D}  and one can verify that the correct relation between compactivity and fluctuations is attained:
 \begin{equation}
\chi^2 \frac{\partial \left< V \right>}{\partial \chi}  =  \left< (V - \left< V \right>)^2\right>   \;\;\;.
\label{e.fluct}
\end{equation}

From this equation, substituting Eqs.  \ref{e.variance} and \ref{e.granT}, we obtain the following relation for the quantity $k$:
 \begin{equation}
k =  \frac{\partial \left< V \right>}{\partial \chi}   \;\;\;,
\label{e.k}
\end{equation}
$k$ measures therefore the amount of volume that must be added to the system in order to increase of one `granular degree' the compactivity.  
The analogous quantity for molecular gasses is: $\partial Energy/\partial Temperature$, which is the \emph{specific heat}.
In analogy with ordinary thermodynamics this `specific heat' is expected to be sensitive to changes in the system's internal properties.
Ultimately, $k$ counts the number of elementary cells, it is therefore a measure of the number of degrees of freedom.

In this section we have shown that the present theory is consistent with all the equilibrium requirements in an ordinary statistical mechanics  approach such as the molecular gas thermodynamics.
Pushing forwards such analogy we can see that if the compactivity is the analogous of the temperature in a thermodynamics system, then Eq.\ref{e.pV1D}  must correspond to the Maxwell-Boltzmann distribution for granular systems.

\section{Theory validation: Vorono\"{\i}  volume distribution}
\label{s.6} \label{s.7}

We have tested the validity of the present theory over a set of packings generated by means of  a modified Lubachevsky-Stillinger algorithm \cite{Lubachevsky90,Donev05b,Skoge06}.
The algorithm starts from random points in space and it makes them uniformly grow into non-overallping hard spheres with the sphere positions evolving in time  according to Newtonian dynamics.
The algorithm uses a cubic box and periodic boundary conditions. 
The simulation is ended when a jammed state with diverging collision rate is reached.
Large expansion rates produce jammed configurations at low packing fractions whereas slower growth rates lead to larger packing fractions. 
For very slow rates, crystalline nuclei with large packing fractions  are formed.  
In our simulations we used 10000 spheres with unitary diameters obtaining  jammed configurations with packing fractions between 0.56 and 0.65 by varying the growth rate between 500 and 0.000001. 
We  also generated non-jammed configurations in the range of packing fractions between 0.1 and 0.55 by keeping the growth rate at 0.001 and arresting the simulation once the desired packing fraction was reached.
We also compare the theoretical predictions with experimental results concerning both acrylic beads in air and glass beads in water from the ANU database on disordered packings \cite{WebGrain}.
Details on these experimental systems are reported in \cite{AsteKioloa,AstePRE05,AsteEPL07}.

Within the framework of the present theoretical approach the packings generated by means of numerical simulations can be considered as prepared with a `single trial'. 
The ergodic set for the jammed packings can be identified with the set of all configurations that can be achieved  with a given growth rate.
For the non-jammed packing configurations the ergodic set can be identified with the ensemble of non-overalpping spheres at a given packing fraction. 
In the case of dry acrylic and glass beads experiments the ergodic set is the set of all mechanically stable packings achievable with the specific preparation action \cite{Aste05rev}. 
Finally, we point out that, in the case of the glass beads in water, the preparation by means of a fluidized bed technique was specifically tuned with a sufficiently large number of repeated flows to ensure that the stationary state was reached \cite{AsteEPL07}.
The fact that in all the above systems the statistical equilibrium is reached is a `working hypothesis' that will be supported, later in this section, by the very good agreement between the observations and the theoretical (equilibrium) predictions.

Equation \ref{e.pV1D} is the main outcome of the present theory and it can be directly tested on the observations from experiments and computer simulations.
The distribution $f(V,k)$ predicted by Eq. \ref{e.pV1D} is valid for \emph{any} aggregate of $k$ elementary cells.
Therefore, the same equation must describe both the volume fluctuations at the level of a single grain as well as the fluctuation at the level of the whole system.

\begin{figure*} 
\centering
{\includegraphics[width=0.95\textwidth]{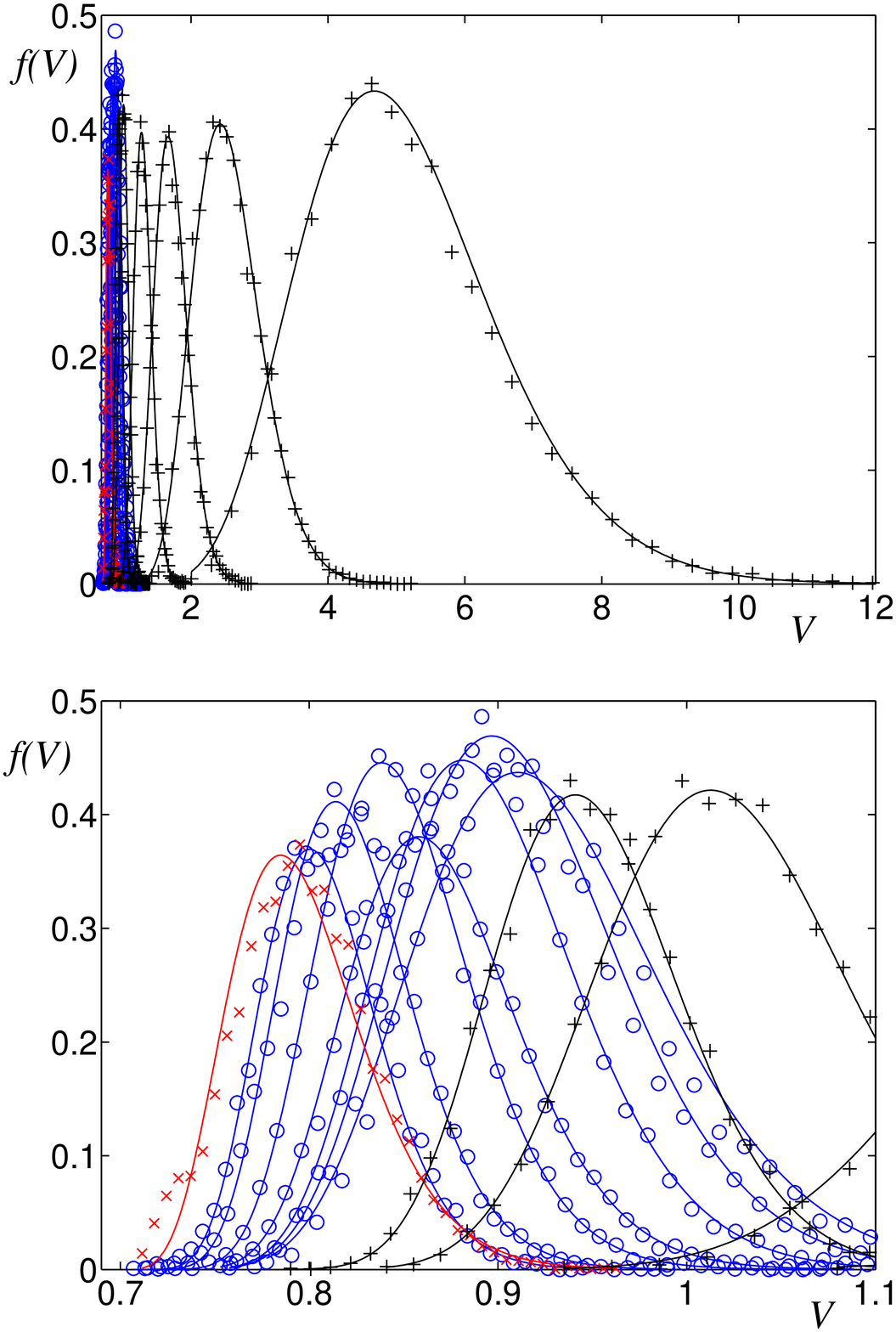}}
\caption{\footnotesize (Color online).  
(top) Vorono\"{\i}  volume distributions from numerically generated equal-sphere packings at packing fractions in a range between 0.1 and 0.65.
Different symbols refer to packings with different properties: jammed ($\circ$ symbols), un-jammed ($+$ symbols) and packing with crystalline inclusions ($\times$ symbols).
The lines are plots of Eq.\ref{e.pV1D} with the parameter $k$ calculated from $k =  (\bar V  -V_{min})^2/\sigma^2_v$ (Eq.\ref{e.variance}).
(bottom) The details of the jammed distributions at packing  fractions between 0.55 and 0.65.}
\label{f.LSvor}
\end{figure*}

\begin{figure*} 
\centering
{\includegraphics[width=0.95\textwidth]{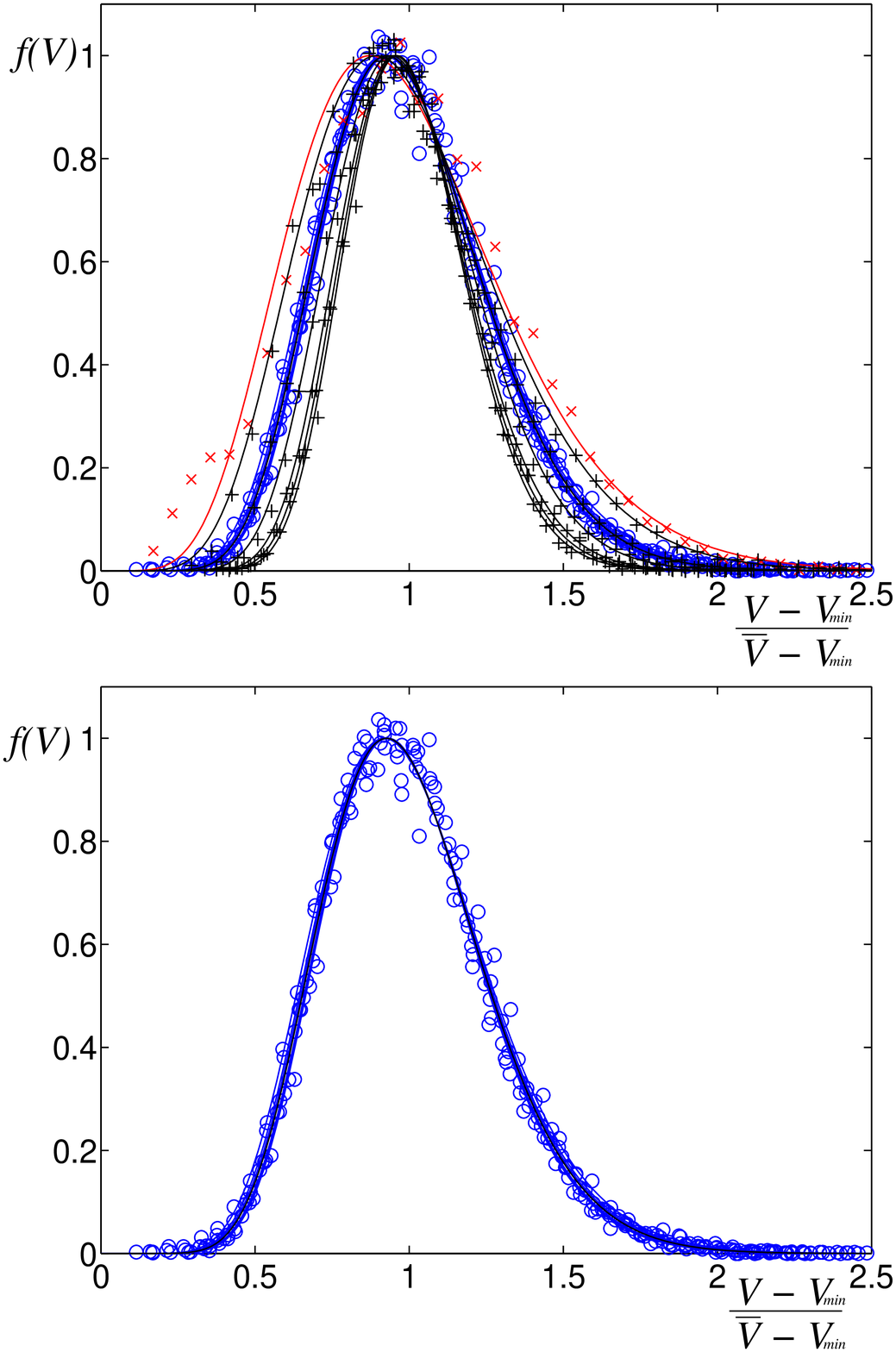}}
\caption{\footnotesize (Color online).  
Vorono\"{\i}  volume distributions  plotted vs. $(V - V_{min})/(\bar V - V_{min})$.
(top) All distributions:  jammed ($\circ$), un-jammed ($+$) and packing with crystalline inclusions ($\times$).
(bottom) The details of the jammed data show that they all collapse onto distributions in good agreement with Eq.\ref{e.pV1D} with very similar values of $k$ (between 12.2 and 14.5).
}.
\label{f.LSvorColl}
\end{figure*}

\subsection{Local volume fluctuations}
Let us first study the system at local level investigating the volume fluctuations at the level of a single grain. 
For this purpose we must first implement a  method to assign a portion of volume associated to each grain in the packing.
A natural way to divide space into cells `warped' around each grain is the Vorono\"{\i} partition where cells are associated to the  portion of space closest to  a grain center  respect to any other centre in the packing.
Such a partition is particularly meaningful in the case of packing of equal spheres because in this case the Vorono\"{\i} cell is always circumscribing the internal sphere.
Extension to grains with more generic shapes and sizes can be implemented.

Figure \ref{f.LSvor} shows the resulting distribution of the Vorono\"{\i} volumes. 
One can see that such distributions span a very broad interval of volumes between 0.69 and 12 with large differences between different samples.
Lines in Fig.\ref{f.LSvor} are the plots of $f(V,k)$ (Eq.\ref{e.pV1D}) with the parameter $k$ calculated from $k =  (\bar V  -V_{min})^2/\sigma^2_v$ (Eq.\ref{e.variance}).
One can see that all the distributions follow well the theoretical prediction.
However, given the broad and different domains in which such distributions are defined, a better visualization of such agreement in necessary.
This can be obtained by an appropriate shift and rescaling of the variable. 
Indeed, from Eq. \ref{e.pV1D}, one can see that distributions characterized by similar values of $k$ must result into similar behaviors when plotted as  function of $(V-V_{min})/(\bar V - V_{min} )$.
Figure \ref{f.LSvorColl} shows the plot of all the distributions in function of such shifted-rescaled variable.
We note that all the distributions from jammed configurations  (bottom figure) collapse into very similar behaviors well described by $f(V,k)$ with $k$ in the range between 12.2 and 14.5.
Un-jammed configurations (top figure) also reveal very good agreement with $f(V,k)$ but in this case the parameter $k$ spans a larger interval  ($9 \le k \le 25$).
We also note that the distribution with crystalline inclusions ($\times$ symbols) deviates slightly from the predicted behavior of $f(V,k)$ revealing the emergence of a second peak at small volumes (corresponding to the close-packed crystalline inclusions).

\begin{figure*} 
\centering
{\includegraphics[width=0.95\textwidth]{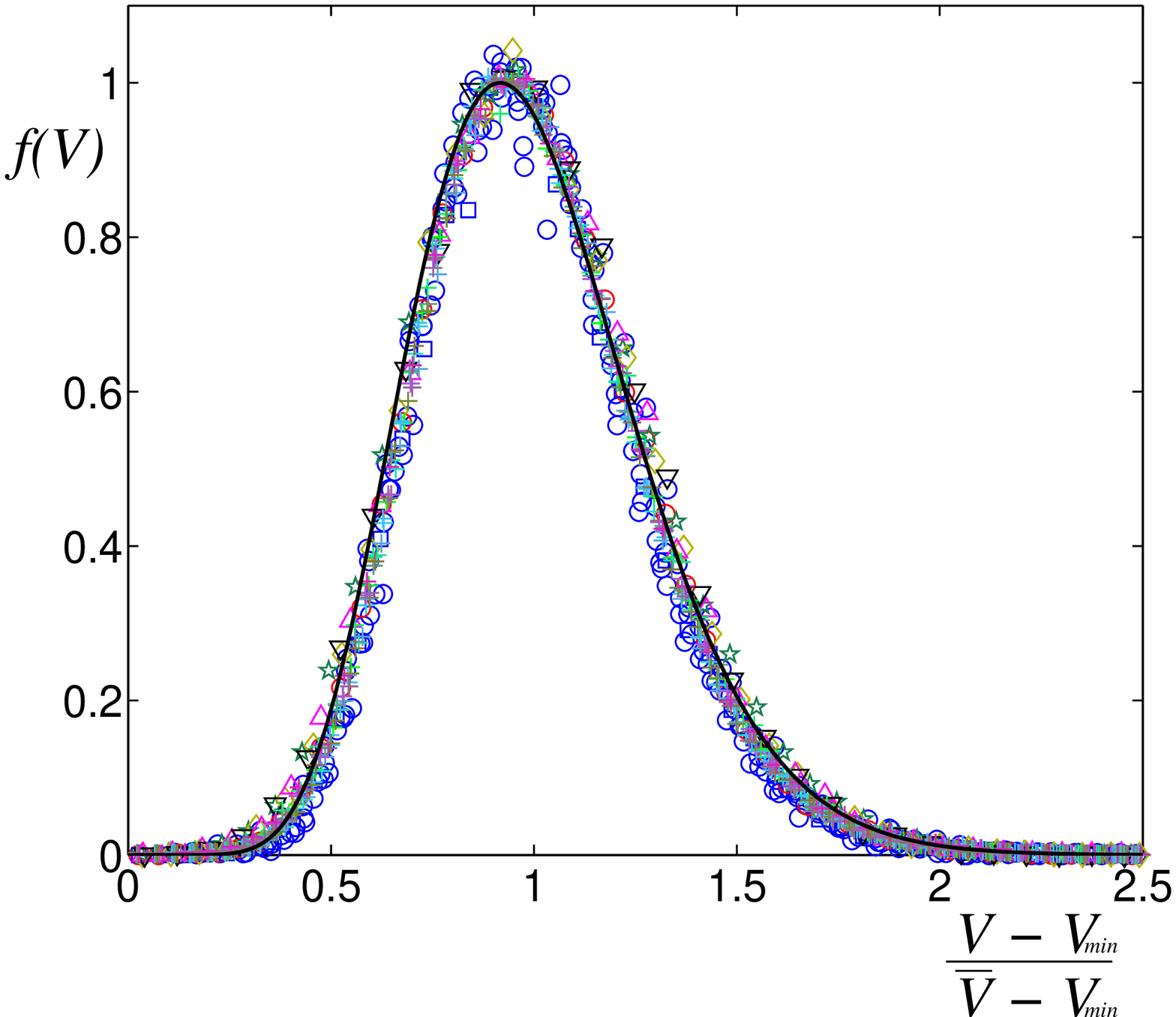}}
\caption{\footnotesize (Color online).  
Comparison between the experimental Vorono\"{\i}  volumes distributions  for all the 18 experimental samples with acrylic and glass beads in air and water \cite{AsteEPL07} and all the numerical jammed configurations  ($\circ$) obtained by using the   modified Lubachevsky-Stillinger algorithm \cite{Lubachevsky90,Donev05b,Skoge06}.
Experiments with dry acrylic beasd in air have symbols:  A ($\lhd$), B ($\Box$), C ($\star$), D ($\diamond$), E ($\bigtriangleup$), F ($\bigtriangledown$).
Experiments with glass beads in water have symbols $+$.
When plotted vs. $(V - V_{min})/(\bar V - V_{min})$ all the data collapse to a common behavior  which is described  well by $f(V,k=12)$ (black line).
}
\label{f.LSvorCollExp}
\end{figure*}

The numerical results analyzed in the previous paragraph refer to rather `ideal' packings made of non overlapping identical spheres.
Remarkably, the same good agreement with $f(V,k)$ (Eq.\ref{e.pV1D}) was also found \cite{AsteEPL07}  in real experimental systems of acrylic spheres in air \cite{AsteKioloa,AstePRE05,Aste05rev,AstePRL06,AsteEPL07} and glass spheres in water  \cite{AsteEPL07} packed at different packing fractions by following different preparation protocols.
Figure \ref{f.LSvorCollExp} shows that all the experimental distributions  and the jammed configurations obtained numerically in the present work, collapse into very similar behaviors  well described by $f(V,k)$ with $k$ in a narrow range around $k \sim 12$.  
We note that, the probability distribution function is well reproduced in all the range of volumes; over eighteen different experimental systems; over eight different computer simulations; over a range of packing fractions between 0.55 to 0.64 and with a statistics performed over more than one million local configurations.

\begin{figure*} 
\centering
{\includegraphics[width=0.95\textwidth]{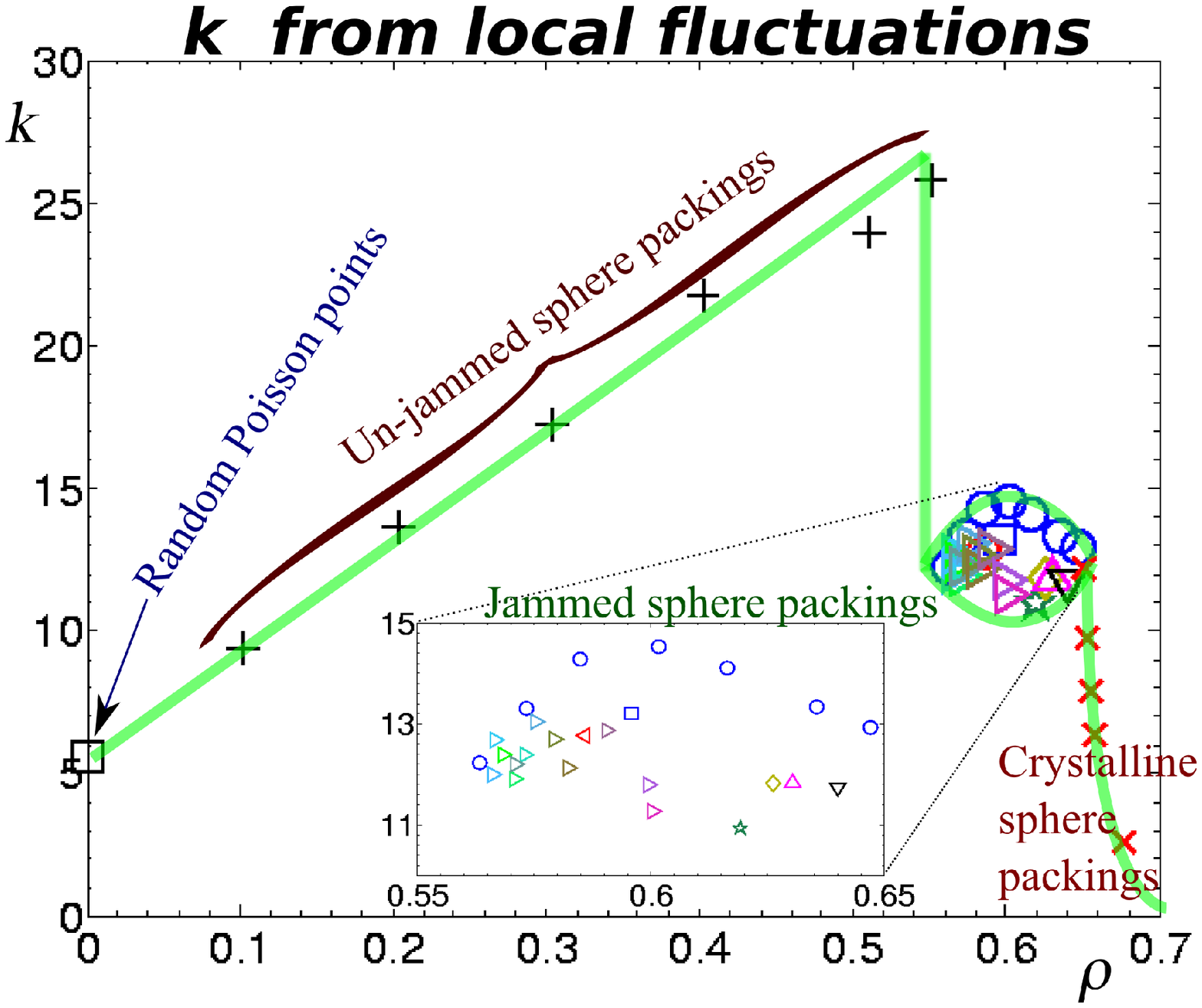}}
\caption{\footnotesize (Color online).  
Behavior of $k$ calculated from $k =  (\bar V  -V_{min})^2/\sigma^2_v$ (Eq.\ref{e.variance}) vs. packing fraction $\rho$.
This plot refers to the statistics on local configurations computed over the fluctuations of the Vorono\"{\i} cells volumes. 
Neat changes in $k$ are observed when spheres first jam ($\rho \sim 0.55$) and when crystalline nuclei start to form ($\rho > 0.65$).
The `$+$' refers to simulations of un-jammed packings of spheres; the `$\circ$' are jammed packings simulated by using different growth rates and `$\times$ is a jammed sample with crystalline inclusions; the symbols `$\rhd$' correspond to 12 experiments with glass spheres in water prepared by means of fluidized beds technique  \cite{AsteEPL07} ; finally, the 6 symbols `$\lhd$, $\Box$, $\star$, $\diamond$, $\bigtriangleup$, $\bigtriangledown$' refer to 6 experiments with acrylic beads in air prepared by means of different methods \cite{AsteKioloa,AstePRE05,Aste05rev,AstePRL06}. 
}
\label{f.krho}
\end{figure*}

The impressive fact of such an agreement  is that these systems are very different (ideal Newtonian spheres, acrylic beads in air and glass beads in water) and they are prepared in very different ways (pouring, tapping, fluid flows, molecular dynamics shearing).
The collapse of all these distributions around $f(V,k\sim 12)$ suggests that there  are universal properties that determine the  packing configurations which are very little sensitive to the preparation method.
Interestingly, we observe that un-jammed configurations are also very well described by $f(V,k)$ but with parameters $k$ which are different from the one of the jammed case.
Figure \ref{f.krho} reports the estimates for the parameter $k$ calculated from the relation $k =  (\bar V  -V_{min})^2/\sigma^2_v$ (Eq.\ref{e.variance}).
The value at zero packing fraction ($k = 5.586$) was calculated analytically for random Poisson points in three dimensions  \cite{Gilbert62,Pineda04}.
Figure \ref{f.krho} shows that, for non jammed configurations,  the value of $k$ increases almost linearly by increasing the packing fraction.
Then the value of the parameter $k$ collapses around 12 when the system gets into a jammed configuration.
The inset in the figure shows that there are sizable differences between different systems and between the same system at different packing fractions.
A careful look shows a rather neat variation in the value of $k$ for the experimental systems around the packing fraction $\rho \sim 0.6$.
Special properties associated to the global volume fluctuations at $\rho \sim 0.6$ was also observed in fluidized bed experiments by Schr\"oter et al.  \cite{Schroder05}.
Above the packing fraction $\sim 0.645$, the packings generate partially crystallized regions and the change in the kind of structural organization is reveled by a sharp drop in the value of $k$ that eventually will tend to zero in the crystalline limit.
We observe that the distribution $f(V,k)$ (Eq.~\ref{e.pV1D}) with $k \sim 12$ is also followed by the  Vorono\"{\i} volumes from simulations of granular packings reported in \cite{Ciamarra07}.
An equivalent collapse for a complete different system concerning for simulations of a polymer melt, water, and silica was reported by by Starr \cite{Starr02} where volume distributions were plotted vs. $(v-\left< v \right>)/\sigma_v$.
It is straightforward to see that such a scaling coincides to the one proposed in this paper, indeed $(v-\left< v \right>)/\sigma = \sqrt{k}((v-v_{min})/(\left< v \right>-v_{min})-1)$.
A careful look at the Starr's data indicates a good agreement with  $f(V,k\sim 20)$.

\begin{figure*} 
\centering
{\includegraphics[width=0.95\textwidth]{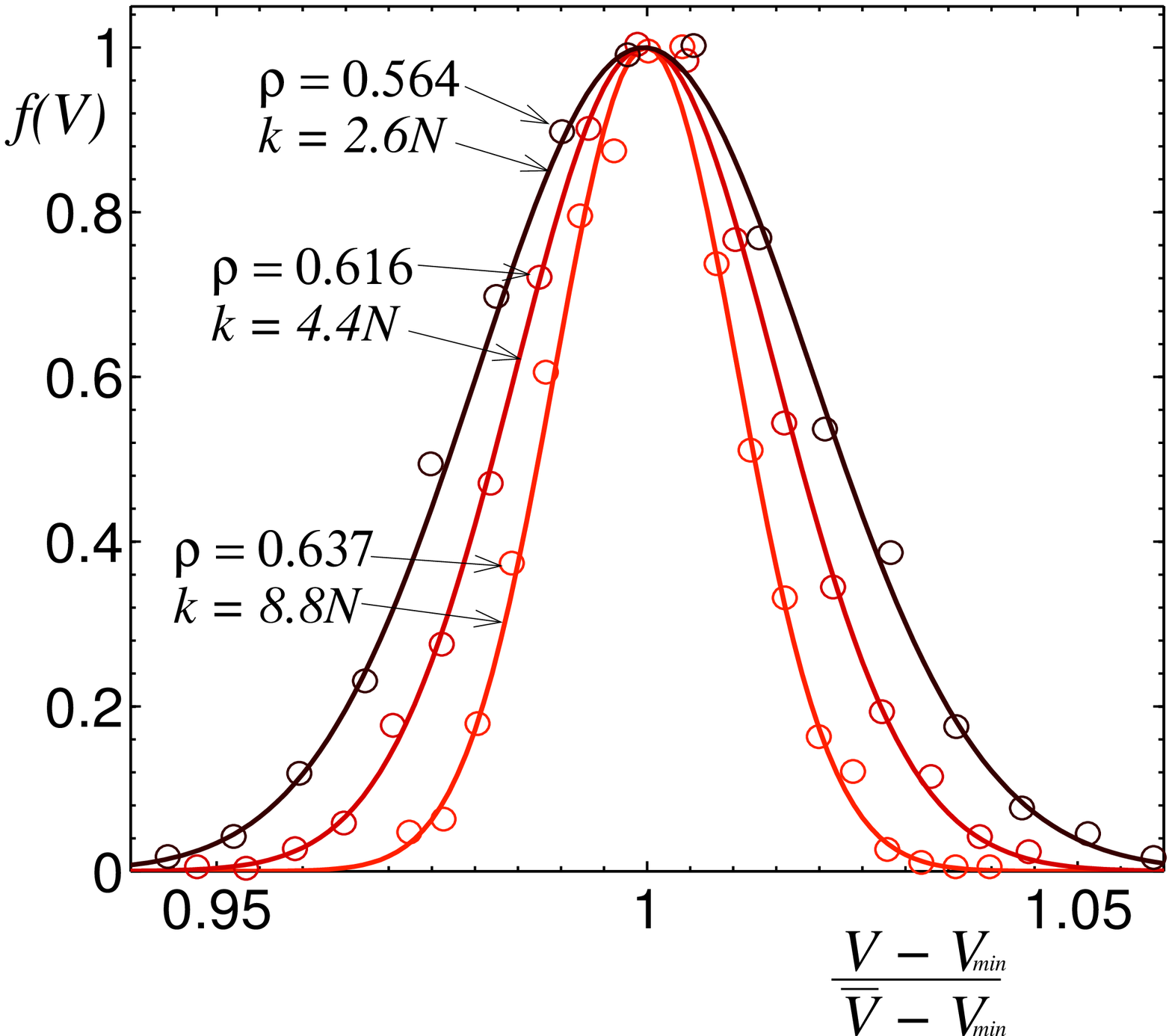}}
\caption{\footnotesize (Color online).  
Fluctuations of the whole sample volume (symbols) from large number of repeated simulations of jammed packings.
The distributions are from samples each containing $n=1000$ spheres and generated with growth rates 0.01 ($\rho = 0.637$), 0.05 ($\rho = 0.616$) and 1000 ($\rho = 0.564$).
The lines are the theoretical prediction from Eq.\ref{e.pV1D}. They use no adjustable parameters: the parameter $k$ is calculated from $k =  (\bar V  -V_{min})^2/\sigma^2_v$ (Eq.\ref{e.variance}) and $V_{min}=N/\sqrt{2}$.
}
\label{f.AgrDist}
\end{figure*}

\begin{figure*} 
\centering
{\includegraphics[width=0.95\textwidth]{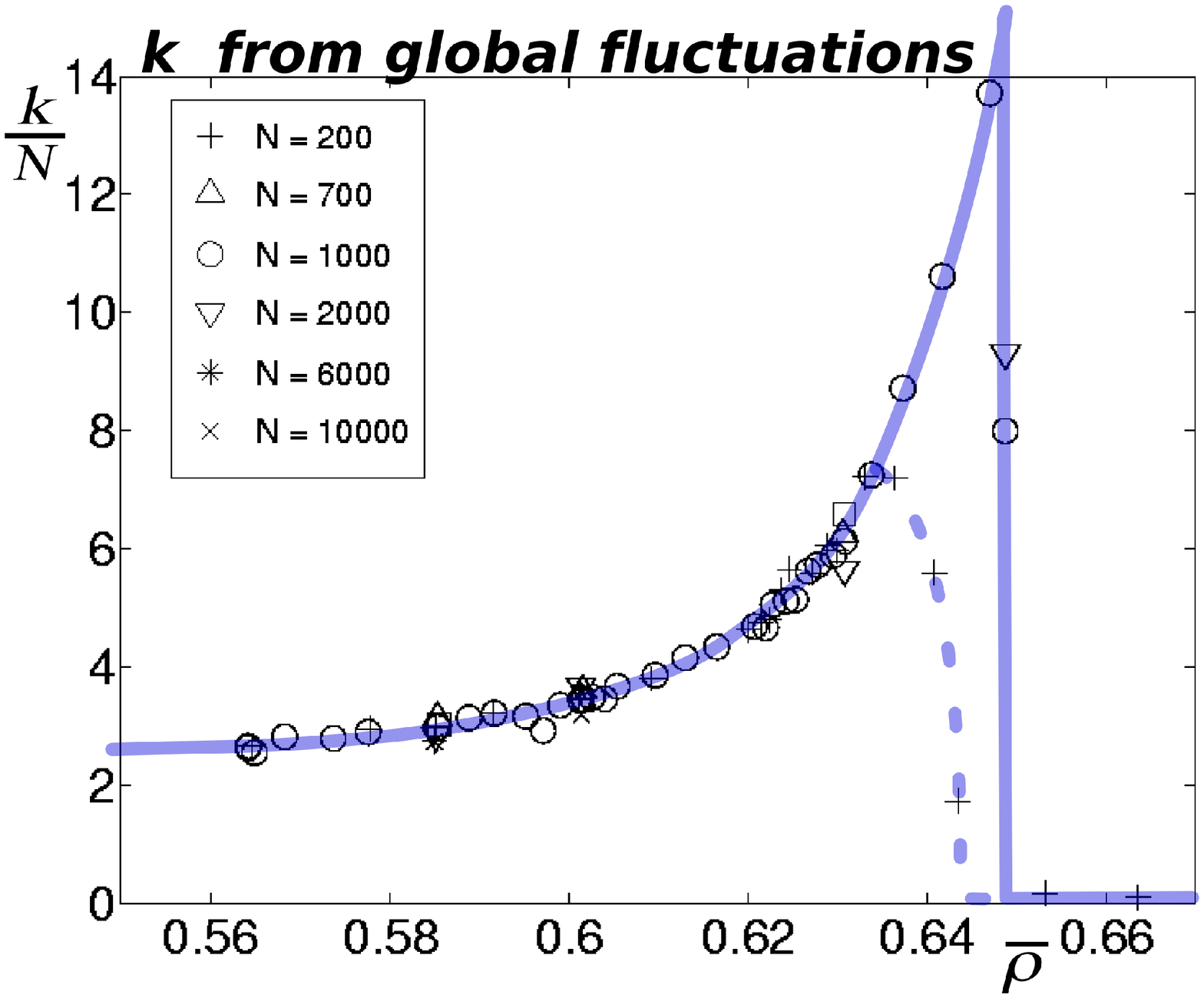}}
\caption{\footnotesize (Color online).  
The parameter $k$ divided by the number of spheres in the sample ($N$) is plotted versus the average packing fraction  $\bar \rho = \pi /(6 \bar V)$ for several samples with total number of spheres between 200 and 10000.
The value of $k$ is computed from $k/N = 1/N (\bar V - V_{min})/ \sigma^2_v =  (\frac{\pi}{6 \bar \rho} - \frac{1}{\sqrt{2}})^2/\sigma^2_v$ over a large set of samples prepared by repeating a large number of simulations using the same growth rate for each average packing fraction.
The full line is a guide for eyes showing the sharp transition occurring around 0.645 for samples with $N\ge 700$, the dotted line highlights that small samples ($N=200$) start the decay at smaller packing fractions around $\rho \simeq 0.635$.
}
\label{f.kNrho}
\end{figure*}

\subsection{Global  volume fluctuations of the whole system}

We have seen in the previous discussion that the volume fluctuations at the level of a single grain are very well described by $f(V,k)$ and we have observed that the value of the parameter $k$ is a good indication of the system state (un-jammed, jammed, poly-crystalline).
Let us now verify if the present theory can also correctly predict the volume fluctuations at the level of the whole sample.
This requires the analysis of the distribution of the volumes of large number of equivalent samples made by means of the same procedure.
Differently from the single-grain case here we can only study jammed configurations. 
We have prepared such samples by using the modified Lubachevsky-Stillinger algorithm \cite{Lubachevsky90,Donev05b,Skoge06}  generating a large number of jammed packings containing different numbers of grains (between $N=200$ to $N=10000$) and prepared at different growth rates (between 0.0005 and 1000).
For each growth rate and size we run at least 200 independent simulations and we measure the total volume occupied by  the jammed spheres.
Three examples of the resulting statistical distributions of volumes are shown in Fig.\ref{f.AgrDist}.
Again we find that  Eq.\ref{e.pV1D} predicts extremely well the observed fluctuations.
The agreement is obtained without using any adjustable parameter.
Indeed, to calculate the parameter $k$ we compute the average volume $\left< V \right> = \bar V$, the variance $\sigma^2_v =  \left< V^2 \right> - \left< V \right>^2$ and we use $k =  (\bar V  -V_{min})^2/\sigma^2_v$ (Eq.\ref{e.variance}).
In this case, $V_{min}$ is fixed to the minimum volume attainable by a packing of $N$ equal spheres with unit diameter, which is the volume occupied by a fcc crystalline configuration: i.e. $N/\sqrt{2}$ (for spheres with unitary diameters). 

One can note that the distribution tends to narrowing  during compaction.
The quantity $k$ is changing with the packing fraction and, for example, in the three examples in Fig.\ref{f.AgrDist} we measure:   $k = 2.6 N$ at  $\rho = 0.564$,  $k = 4.4 N$ at $\rho = 0.616$ and $k = 8.8 N$ at $\rho = 0.637$.
More generally we compute the average volume and the variance for a very large set of data over a range of packing fractions within 0.55 and 0.66.
The resulting values of the parameter $k$ rescaled by the number of spheres in each sample $N$ are plotted in Fig.\ref{f.kNrho}.
We note that the data from samples with various sizes (between $N= 200$ to $N= 10000$) all fall on similar values when rescaled by $N$ indicating that the parameter $k$ scales linearly with the number of spheres.
The rescaled parameter $k/N$ takes values a little below 3 at the loose packing limit ($\rho \sim 0.55$).
Then the value of the parameter grows consistently when the packing densifies and it reaches a peak at the random close packing limit $\rho \sim 0.64$.
Afterwards, $k/N$ drops sharply and eventually goes to zero when the packing becomes crystalline.
The small samples with  $200$ spheres show a slightly different behavior with the peak occurring before $\rho \sim 0.64$.
We verified that this is a finite-size effect  due to the formation of some poly-crystalline samples.

It is therefore clear that the quantity $k$ is very sensitive to the changes in the system's internal structure and the peak at the random close packing limit indicates that the packing must reorganize in order to proceed with compaction. 
A similar discontinuity at both the loose and closed packing limits was observed in the study of $k$ from the local Vorono\"{\i} volumes statistics (Fig.\ref{f.krho}). 
However, the values of the local quantity ($k$) and the rescaled global quantity ($k/N$) are different.
This should not be of any surprise.
Indeed, $k$ counts the number of elementary cells contributing to the system's volume. 
But, some elementary cells are sheared among neighboring Vorono\"{\i} cells, and therefore the number that contributes to the volume of $N$ Vorono\"{\i} cells is not necessary $N$ times the number that contributes to one single cell.
Furthermore, it is known that volume-correlations play an important and unusual role in these systems \cite{AsteScaling07}.
It has been shown both in computer simulations and experiments \cite{Torquato03,Donev05c,AsteScaling07} that the scaling of the volume fluctuations is not a simple linear function of the portion of sample investigated and there are evidences of a dependence on the preparation procedure and on the packing fraction. 
Although not surprising, such different values and behaviors of the local and global $k$ have important implications.
For instance this implies that the granular temperature measured at local level is different from the one measured from the global fluctuations.
In this respect the `proper' granular temperature is the one associated to the global volume fluctuations which we have shown is an intensive quantity since $k$ is scaling linearly with $N$ (at least for samples with more than 200 spheres, Fig.\ref{f.kNrho}).   
The `temperature' associated with the local measure of $k$ is an `effective granular temperature'.
The fact that they predict so accurately the volume fluctuations both at local and at global level indicates that the effect of correlations and local geometrical constraints can be treated within a statistical mechanics framework by simply tuning the value of $k$.

\section{Conclusions}
\label{s.9}

In this paper we have shown that a statistical mechanics theory for granular matter can be always deductively constructed from two simple physical assumptions: 
1) the microscopic system's structure can be completely described in terms of a finite set of quantities;
2) after performing a given experiment, the probability to find the system in a certain static state depends only on the previous state (and on the experiment itself).
These two simple assumptions lead to a complete  statistical mechanics description for these systems where both transient and equilibrium properties can be treated. 
We show that an expression for the volume fluctuations can be explicitly obtained (Eq.\ref{e.ME}) by introducing the following three additional assumptions: 
i) the microscopic state can be encoded in terms of the properties of  a set of local cells with volumes $v_i$; 
ii) the cells properties are either fully described by their volumes $v_i$ or they are independent from $v_i$; 
iii) any space partition with cells of volumes $v_i \ge v_{min}$ satisfying the condition $\sum_i v_i = V$   corresponds to an attainable packing.
A comparison with a large number of experiments and computer simulations shows remarkably good agreements with the theoretical predictions in a wide range of packing fractions and for several differently-prepared systems (Section \ref{s.6} and Figs.  \ref{f.LSvor}, \ref{f.LSvorColl} and \ref{f.LSvorCollExp}).

We have identified a quantity, $k$, which is very sensitive to the system's structural organization.
It results  that such quantity is the equivalent of the specific heat in ordinary thermodynamics.
We have demonstrated that $k$ can be easily computed empirically (Eq.\ref{e.variance}) by measuring the average volume and the variance in a set of trials.
A large number of measures from both the local statistics at the level of the single grain and at the global level of the whole sample show the signatures of clear structural transitions marked by peaks and discontinuities in the values of $k$ occurring at both the random loose packing and at the random close packing limits (Figs.\ref{f.krho} and \ref{f.kNrho}).
Similar peaks in the specific heat are observed at phase transitions in thermodynamic systems.
Understanding whether proper thermodynamical phase transitions are occurring at the he random loose packing limit and the random close packing limit is the subject of current investigations.

We have demonstrated that the present theory can predict both the volume fluctuation at the level of a single grain (Vorono\"{\i} cells) and at the level of the whole system (see Figs.  \ref{f.LSvorCollExp} and \ref{f.AgrDist}).
The remarkable quantitative agreement between theory, experiments and simulations at both local and global levels is very encouraging demonstrating that statistical mechanics is a powerful tool to study and predict the properties of complex materials.
We have discussed that the local effects of correlations and geometrical constraints can be taken into account by using an `effective' local granular temperature.
Further work must be devoted to better understand the link between local and global properties and identify the length-scale above which the system becomes extensive.
An important test to establish the correct granular temperature might use an independent measure from the fluctuation-dissipation theorem as discussed in \cite{Barrat00,Makse02}. 
The use of non extensive forms of entropy instead of the classical Gibb's entropy might be right approach to formulate a theory which consistently describes both local and global levels in terms of a unified generalized granular temperature.

\subsection*{Acknowledgements}
Many thanks to Enrico Scalas who has first pointed out the relevance of the Penrose's approach  to complex systems.
We thank  Antonio Coniglio, Mario Nicodemi, Massimo Pica Ciamarra, Matthias Schr\"oter and  Harry Swinney for helpful discussions.
Many thanks to T.J. Senden, M. Saadatfar, A. Sakellariou, A. Sheppard, A. Limaye for the help with tomographic data.
This work was partially supported by the ARC discovery project DP0450292.


\end{document}